\documentclass[12pt]{article}
\usepackage{latexsym}
\usepackage{amsmath}
\usepackage{amssymb}
\usepackage{amsfonts}
\usepackage{comment}
\usepackage{graphicx}
\usepackage{verbatim}
\newcommand{\be}{\begin{equation}}
\newcommand{\ee}{\end{equation}}
\newcommand{\bea}{\begin{eqnarray}}
\newcommand{\eea}{\end{eqnarray}}
\newcommand{\mn}{{\mu\nu}}
\newcommand{\mpl}{m_{\rm pl}}
\newcommand{\lpl}{L_{\rm pl}}

\newcommand{\calh}{{\mathcal H}}

\newcommand{\dd}{{\rm d}}

\newcommand{\tildeomega}{{\widetilde{\omega}}}

\newcommand{\ghs}{{\Theta}} 
\newcommand{\drt}{\widetilde{\delta\rho}}
\newcommand{\htilde}{\widetilde{H}}
\newcommand{\eq}{{\rm eq}}

\begin{document}
\baselineskip=20pt

\begin{flushright} CALT 68-2797  \end{flushright}

\begin{center}
{\Large{\bf Unitary Evolution and Cosmological Fine-Tuning}}

\vspace*{0.1in}
Sean M.\ Carroll and Heywood Tam

\it California Institute of Technology\\ {\tt seancarroll@gmail.com, heywood.tam@gmail.com}
\vspace*{0.1in}
\end{center}

\begin{abstract}
Inflationary cosmology attempts to provide a natural explanation 
for the flatness and homogeneity of the observable universe. 
In the context of reversible (unitary) evolution, this goal is difficult to satisfy, as
Liouville's theorem implies that no dynamical process can 
evolve a large number of initial states into a small number of final states.  
We use the invariant measure on solutions to Einstein's equation to quantify the
problems of cosmological fine-tuning.  
The most natural interpretation of the measure
is the flatness problem does not exist; almost all Robertson-Walker cosmologies 
are spatially flat.  The homogeneity of the early universe, however,
does represent a substantial fine-tuning; the horizon problem is real.
When perturbations are taken into account, inflation only occurs in a negligibly
small fraction of cosmological histories, less than $10^{-6.6\times 10^7}$.
We argue that while inflation does not affect 
the number of initial conditions that evolve into a late universe like our own, 
it nevertheless provides an appealing target for true theories of initial conditions,
by allowing for small patches of space with sub-Planckian curvature to grow
into reasonable universes. 
\end{abstract}

\baselineskip=14pt

\newpage

\tableofcontents

\newpage

\section{Introduction}

Inflationary cosmology \cite{Guth:1980zm,Linde:1981mu,Albrecht:1982wi} has come
to play a central role in our modern understanding of the universe.
Long understood as a solution to the horizon and flatness problems, the success
of inflation-like perturbations (adiabatic, Gaussian, approximately scale-invariant)
at explaining a multitude of observations has led most cosmologists to believe
that some implementation of inflation is likely to be responsible for determining the initial
conditions of our observable universe.

Nevertheless, our understanding of the fundamental workings of inflation 
lags behind our progress in observational cosmology.  Although there are many models,
we do not have a single standout candidate for a specific particle-physics realization
of the inflaton and its dynamics.  The fact that the scale of inflation is likely to be
near the Planck scale opens the door to a number of unanticipated physical
phenomena.  Less often emphasized is our tenuous grip
on the deep question of whether inflation actually delivers on its promise:
providing a dynamical mechanism that turns a wide variety of plausible
initial states into the apparently finely-tuned conditions characteristic of our
observable universe.  

The point of inflation is to make the evolution of our observable universe seem natural.    
One can take the attitude that initial conditions are simply to be accepted, rather than explained -- 
we only have one universe, and should learn to deal with it, rather than seek explanations
for the particular state in which we find it.  In that case, there would
never be any reason to contemplate inflation.  The reason why inflation seems 
compelling is because we are more ambitious:  we would like to understand
why the universe seems to be one way, rather than some other way.  By its own
standards, the inflationary paradigm bears the burden of establishing that 
inflation is itself natural (or at least more natural than the alternatives).

It has been recognized for some time that there is tension between this goal and
the underlying structure of classical mechanics (or quantum mechanics, for
that matter).  A key feature of classical mechanics is conservation of
information: the time-evolution
map from states at one time to states at some later time is invertible and volume-preserving,
so that the earlier states can be unambiguously recovered from the later states.
This property is encapsulated by Liouville's theorem, which states that a distribution
function in the space of states remains constant along trajectories; roughly speaking,
a certain number of states at one time always evolves into precisely the same number
of states at any other time.  In quantum mechanics, an analogous property is 
guaranteed by unitarity of the time-evolution operator; most of our analysis here will
be purely classical, but we will refer to the conservation of the number of states as
``unitarity'' for convenience.

The conflict with the philosophy of inflation is clear.  Inflation attempts to account for
the apparent fine-tuning of our early universe by offering a mechanism by which a
relatively natural early condition will robustly evolve into an apparently finely-tuned
later condition.  But if that evolution is unitary, it is impossible for any mechanism to evolve a 
large number of states into a small number, so the number of initial conditions corresponding
to inflation must be correspondingly small, calling into question their status
as ``relatively natural.''  This point has been emphasized by Penrose \cite{penrose},
and has been subsequently discussed elsewhere
\cite{Gibbons:1986xk,Hawking:1987bi,Coule:1994gd,Unruh:1996sf,Hollands:2002yb,Kofman:2002cj,
Hollands:2002xi,Carroll:2004pn,Gibbons:2006pa,Linde:2007fr}.   
As long as it operates within the framework of unitary evolution, the best inflation can
do is to move the set of initial conditions that creates a smooth, flat universe at late times
from one part of phase space to another part; it cannot increase the size of that 
set.

As a logical possibility, the true evolution of the universe may be 
non-unitary.  Indeed, discussions of cosmology often proceed as if this were the
case, as we discuss below. The justification for this perspective is that a comoving 
patch of space is smaller at earlier times, and therefore can accommodate fewer
modes of quantum fields.  But there is nothing in quantum field theory, or anything
we know about gravity, to indicate that evolution is fundamentally non-unitary.
The simplest resolution is to imagine that there are a large number of states that
are not described by quantum fields in a smooth background ({\it e.g.}, with
Planckian spacetime curvature or the quantum-mechanical version thereof).  
Even if we don't have a straightforward description of the complete set of
such states, the underlying principle of unitarity is sufficient to imply that they
must exist.  

If unitary evolution is respected, there is nothing special about ``initial'' states; the state
at any one moment of time specifies the evolution just as well as the state at any other
time.  In that light, the issue of cosmological fine-tuning is a question about 
\emph{histories}, not simply about initial conditions.
Our goal should not be to show that generic initial conditions give rise to the early universe
we observe; Liouville's theorem forbids it.  Given the degrees of freedom
constituting our observable universe, and the macroscopic features of their current
state, the vast majority of possible evolutions do not arise from a smooth Big Bang beginning.
Therefore, a legitimate explanation for cosmological fine-tuning would show that not all
histories are equally likely -- that the history we observe is very natural
within the actual evolution of the universe, even though it belongs to a tiny fraction of all
conceivable trajectories.  In particular, a convincing scenario would possess the property
that when the degrees of freedom associated with our observable
universe are in the kind of state we currently find them in, it is most often in the
aftermath of a smooth Big Bang.

We can imagine two routes to this goal:  either our present condition only occurs once,
and the particular history of our universe is simply highly non-generic (perhaps due to an 
underlying principle that determines the wave function of
the universe); or conditions like those of our observable universe occur many times within
a much larger multiverse, and the dynamics has the property that most appearances of our
local conditions (in some appropriate measure) are associated with smooth Big-Bang-like
beginnings.  In either case, inflation might very well play a crucial role in the evolution of 
the universe, but it does not by itself constitute an answer to the puzzle of cosmological
fine-tuning.

In this paper we try to quantify the issues of cosmological fine-tuning in the context of unitary
evolution, using the canonical measure on the space of 
solutions to Einstein's equations developed by Gibbons, Hawking, and Stewart \cite{Gibbons:1986xk}.  
Considering first the measure on purely Robertson-Walker cosmologies (without
perturbations) as a function of spatial curvature, there is a divergence at
zero curvature.  In other words, curved RW cosmologies are a set of measure
zero -- the flatness problem, as conventionally understood, does not exist.  
This divergence has no immediate physical relevance, as the real world is not described by
a perfectly Robertson-Walker metric.  Nevertheless, it serves as a cautionary
example for the importance of considering the space of initial conditions in a
mathematically rigorous way, rather than relying on our intuition.  We therefore
perform a similar analysis for the case of perturbed universes, to verify that there
is not any hidden divergence at perfect homogeneity.  We find that there is not;
any individual perturbation can be written as an oscillator with a time-dependent
mass, and the measure is flat in the usual space of coordinate and momentum.
The homogeneity of the universe represents a true fine tuning; there is no
reason for the universe to be smooth. 

We also use the canonical measure to investigate the likelihood of inflation.  In the
minisuperspace approximation, we find that inflation can be very probable, depending on the
inflaton potential considered.  However, this approximation is wildly inappropriate for this
problem; it is essential to consider perturbations.  If we restrict ourselves to universes that
look realistic at the epoch of matter-radiation equality, we find that only a negligible fraction
were sufficiently smooth at early times to allow for inflation.  This simply reflects the aforementioned
fact that there are many more inhomogeneous states at early times than smooth ones.

We are not suggesting that inflation plays no role in cosmological dynamics; only that it 
is not sufficient to explain how our observed early universe arose from generic initial conditions.
Inflation requires very specific conditions to occur -- 
a patch of space dominated by potential energy over a
region larger than the corresponding Hubble length \cite{Vachaspati:1998dy} -- 
and these conditions are an extremely small fraction of all possible states.
However, while inflationary conditions are very few, there is
something simple and compelling about them.  Without inflation, when the Hubble parameter
was of order the Planck scale our universe needed to be smooth over a length scale
many orders of magnitude larger than the Planck length.  With inflation, by contrast, 
a smooth volume of order the Planck length can evolve into our entire observable universe.
There are fewer such states than those required by conventional Big Bang cosmology,
but it is not hard to imagine that they are somehow easier to create.  
In other words, given that the history of our observable universe seems non-generic by any
conceivable measure, it seems very plausible that some hypothetical theory of initial
conditions (or multiverse dynamics) creates the necessary initial conditions through the
mechanism of inflation, rather than by creating a radiation-dominated Big Bang universe
directly.  We argue that this is the best way to understand the role of inflation, rather than as a 
solution to the horizon and flatness problems.

The lesson of our investigation is that the state of the universe does appear 
unnatural from the point of view of the canonical measure on the space of
trajectories, and that no choice of unitary evolution can alleviate that 
fine-tuning, whether it be inflation or any other mechanism.
Inflation can alter the set of initial conditions that leads to a universe like ours, but
it cannot make it any larger.  Inflation does not remove the need for a theory of
initial conditions; it brings that need into sharper focus.

\section{The Evolution of our Comoving Patch}

For many years, the paradigm for fundamental physics has been
information-conserving dynamical laws applied to initial data.  A consequence of
information conservation is reversibility:  the state of the system at any one time is
sufficient to recover its initial state, or indeed any state in the past or future.
The goal of this section is to lay out the motivations for treating the degrees of freedom
of our observable universe as a system obeying reversible dynamics, and to 
establish the limitations of that approach.

Both quantum mechanics and classical mechanics feature this kind of unitary
evolution.\footnote{The collapse of the wave function in quantum mechanics is an
apparent exception.  We will not address this phenomenon, implicitly assuming 
something like the many-worlds interpretation, in which wave function collapse is 
only apparent and the true evolution is perfectly unitary.}  In the Hamiltonian
formulation of classical mechanics, a state is an element of phase space,
specified by coordinates $q^i(t)$ and
momenta $p_i(t)$. Time evolution is governed by Hamilton's equations,
\be
  \dot{q}^i = \frac{\partial \calh}{\partial p_i}\, , \quad 
  \dot{p}_i = -\frac{\partial \calh}{\partial q^i}\, ,  
  \label{hamilton}
\ee
where $\calh$ is the Hamiltonian.  In quantum mechanics, a state is given by a
wave function $|\psi(t)\rangle$ which defines a ray in Hilbert space. Time
evolution is governed by the Schr\"odinger equation, 
\be
  \hat{\calh}|\Psi\rangle = i\partial_t|\Psi\rangle,
  \label{schrodinger}
\ee 
where $\hat{\calh}$ is the Hamiltonian operator, or equivalently by the von~Neumann
equation,
\be
  \partial_t\hat\rho = -i[\hat{\calh}, \hat\rho]\,,
  \label{vonneumann}
\ee
where $\hat\rho(t) = |\psi(t)\rangle\langle\psi(t)|$ is the density operator.
In either formalism, knowledge of the state at any one moment of time is sufficient
(given the Hamiltonian) to determine the state at all other times.  While we don't yet
know the complete laws of fundamental physics, the most conservative assumption
we could make would be to preserve the concept of unitarity.  Even without knowing
the Hamiltonian or the space of states, we will see that the principle of unitarity alone
offers important insights into cosmological fine-tuning problems.

Although the assumption of unitary evolution seems like a mild one, 
there are challenges to applying the idea directly to an expanding universe.
We can only observe a finite part of the universe, and the
physical size of that part changes with time.  The former feature implies that
the region we observe is not a truly closed system, and the latter implies that the
set of field modes within this region is not fixed.  Both aspects could be taken to imply
that, even if the underlying laws of fundamental physics are perfectly unitary, it
would nevertheless be inappropriate to apply the principle of unitarity to the the
part of the universe we can observe.

We will take the stance that it is nevertheless sensible to proceed under
the assumption that the degrees of freedom describing our observable universe
evolve according to unitary dynamical laws, even if that assumption is an
approximation.  In this section we offer the justification for this assumption.
In particular we discuss two separate parts to this claim:  that the observable universe
evolves autonomously (as a closed system), and that this autonomous evolution
is governed by unitary laws.

\subsection{Autonomy}

We live in an expanding universe that is 
approximately homogeneous and isotropic on large scales.   We can therefore consider
our universe as a perturbation of an exactly homogenous and isotropic 
(Robertson-Walker) background
spacetime.  Defining a particular map from the background to our physical spacetime
involves a choice of gauge.  Nothing that we are going to do depends on how that
gauge is chosen, as long as it is defined consistently throughout the history of the 
universe.  Henceforth we assume that we've chosen a gauge.

The map from the RW background spacetime to our universe provides two crucial
elements:  a foliation into time slices, and a congruence of comoving geodesic
worldlines.  The time slicing allows us to think of the universe as a fixed set of
degrees of freedom evolving through time, obeying Hamilton's equations.  At each
moment in time there exists an exact value of the (background) Hubble parameter and
all other cosmological parameters.  

\begin{figure}[t]
\begin{center}
\includegraphics[width=.9\textwidth]{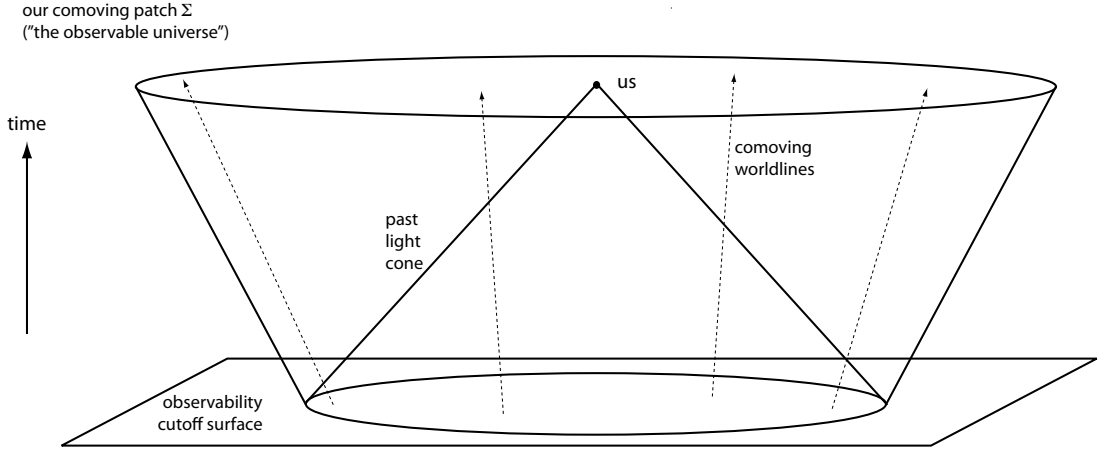}
\end{center}
\caption{The physical system corresponding to our observable universe.  Our comoving
patch is defined by the interior of the intersection of our past light cone with a cutoff 
surface, for example the surface of last scattering.
This illustration is not geometrically faithful, as the expansion
is not linear in time.  Despite the change in physical size, we assume that the space of states is of 
equal size at every moment.}
\label{comoving-patch}
\end{figure}

The notion of comoving worldlines, orthogonal to spacelike hypersurfaces of constant
Hubble parameter, allows us to define what we mean by our comoving patch.
If there is a Big Bang
singularity in our past, there is a corresponding particle horizon, defined by the
intersection of our past light cone with the singularity.  However, independent of the
precise nature of the Big Bang, there is an effective limit to our ability to observe
the past; in practice this is provided by the surface of last scattering, although in principle
observations of gravitational waves or other particles could extend the surface
backwards.  The precise details of where we draw the surface aren't important to our
arguments.  What matters is that there exists a well-defined region of three-space
interior to the intersection of our past light cone with the observability surface past which
we can't see.  Our comoving patch, $\Sigma$, is simply the physical system defined by the
extension of that region forward in time via comoving worldlines, as shown in
Figure~\ref{comoving-patch}.

Our assumption is that this comoving patch can be considered as a set of degrees
of freedom evolving autonomously through time, free of influence from the rest of the
universe.  This is clearly an approximation, as an observer stationed close to the boundary
of our patch would see particles pass both into and out of that region; our comoving
patch isn't truly a closed system.  However, 
the fact that the observable universe is homogenous implies that the net effect of
that exchange of particles is very small.  In particular, we generally don't believe that what
happens inside our observable universe depends in any significant way on what
happens outside.

Note that we are not necessarily assuming that our observable universe is in a pure
quantum state, free of entanglement with external degrees of freedom; such 
entanglements don't affect the local dynamics of the internal degrees of freedom, 
and therefore are complete compatible with the von~Neumann equation
(\ref{vonneumann}).  We are, however, assuming that the appropriate Hamiltonian
is local in space.  Holography implies that this is not likely to be strictly true, but it seems
like an effective approximation for the universe we observe.

\subsection{Unitarity}

Autonomy implies that we can consider our comoving patch as a fixed set of degrees
of freedom, evolving through time.  Our other crucial assumption is that this evolution
is unitary (reversible).  Even if the underlying fundamental laws of physics are unitary,
it is not completely obvious that the effective evolution of our comoving patch evolves
this way.  Indeed, this issue is at the heart of the disagreement between those who 
have emphasized the amount of fine-tuning required by inflationary initial conditions
\cite{penrose,Unruh:1996sf,Hollands:2002yb,Hollands:2002xi,Carroll:2004pn}
and those who have argued that they are natural \cite{Kofman:2002cj,Linde:2007fr}. 

The issue revolves around the time-dependent nature of the cutoff on modes of a
quantum field in an expanding universe.  Since we are working in a comoving patch,
there is a natural infrared cutoff given by the size of the patch, a length scale of order
$\lambda_{\rm IR} \sim a H_0^{-1}$, where $a$ is the scale factor (normalized to unity
today) and $H_0$ is the current Hubble parameter.  But there is also a fixed ultraviolet
cutoff at the Planck length, $\lambda_{\rm UV} \sim \lpl = \sqrt{8\pi G}$.
Clearly the total number of modes that fit in between these two cutoffs increases
with time as the universe expands.  It is therefore tempting to conclude that
the space of states is getting larger.

We can't definitively address this question in the absence of a theory of quantum 
gravity, but for purposes of this paper we will assume that the space of states is
\emph{not} getting larger -- which would violate the assumption of unitarity -- but the
nature of the states is changing.  In particular, the subset of states that can usefully
be described in terms of quantum fields on a smooth spacetime background is
changing, but those are only a (very small) minority of all possible states.

The justification for this view comes from the assumed reversibility of the underlying
laws.  Consider the macrostate of our universe today -- the set of all microstates
compatible with the macroscopic configuration we observe.  For any given amount
of energy density, there are two solutions to the Friedmann equation, one with
positive expansion rate and one with negative expansion rate (unless the expansion
rate is precisely zero, when the solution is unique).  So there are an equal number
of microstates that are similar to our current configuration, except that the universe
is contracting rather than expanding.  As the universe contracts, each of those states
must evolve into some unique state a fixed time later; therefore, the number of states accessible
to the universe for different values of the Hubble parameter (or different moments in
time) is constant.  

Most of the states available when the universe is smaller, however, are not
described by quantum fields on a smooth background.  This is reflected in the
fact that spatial inhomogeneities would be generically expected to grow, rather than
shrink, as the universe contracted.  The effect of gravity on the state counting becomes
significant, and in particular we would expect copious production of black holes.
These would appear as white holes in the time-reversed expanding description.
Therefore, the overwhelming majority of states at early times that could evolve
into something like our current observable universe are not relatively smooth spacetimes
with gently fluctuating quantum fields; they are expected to be wildly inhomogeneous,
filled with white holes or at least Planck-scale curvatures.  

We do not know enough about quantum gravity to explicitly enumerate these states,
although some attempts to describe them have been made (see {\it e.g.} \cite{Mathur:2008ez}).
But we don't need to know how to describe them; the underlying assumption
of unitarity implies that they are there, whether we can describe them or not.  (Similarly,
the Bekenstein-Hawking entropy formula is conventionally taken to imply a large number
of states for macroscopic black holes, even if there is no general description for what
those individual states are.)  

This argument is not new, and it is often stated in terms of the entropy of our comoving
patch \cite{penrose,Carroll:2004pn}.  In the current universe, this entropy is dominated by black holes, 
and has a value of order $S_\Sigma(t_0) \sim 10^{104}$ \cite{Egan:2009yy}.  If all the matter 
were part of a single black hole it would be as large as $S_\Sigma({\rm BH}) \sim 10^{122}$.
At early times, when inhomogeneities were small and local gravitational effects were negligible,
the entropy was of order $S_\Sigma({\rm RD}) \sim 10^{88}$.  If we assume that the
entropy is the logarithm of the number of macroscopically indistinguishable microstates,
and that every microstate within the current macrostate corresponds to a unique 
predecessor at earlier times,
it is clear that the vast majority of states from which our present universe might have 
evolved don't look anything like the smooth radiation-dominated configuration we actually
believe existed (since $\exp[10^{104}] \gg \exp[10^{88}] $).

This distinction between the number of states implied by the assumption of unitarity
and the number of states that could reasonably be described by quantum fields on a smooth
background is absolutely crucial for the question of how finely-tuned are the conditions
necessary to begin inflation.  If we were to start with a configuration of small size and
very high density, and consider only those states described by field theory, we would
dramatically undercount the total number of states.  Unitarity 
could possibly be violated in an ultimate theory, but we will accept it for
the remainder of this paper.

\section{The Canonical Measure}
\label{measureintro}

In order to quantify the issue of fine-tuning in the context of unitary evolution,
we review the canonical measure on the space of trajectories, as
examined by Gibbons, Hawking, and Stewart \cite{Gibbons:1986xk}.
Despite subtleties associated with coordinate
invariance, GR can be cast as a conventional Hamiltonian system, with an infinite-dimensional
phase space and a set of constraints.  The state of a classical system is described
by a point $\gamma$ in a phase space $\Gamma$, with canonical coordinates
$q^i$ and momenta $p_i$.  The index $i$ goes from $1$ to $n$, so that phase space
is $2n$-dimensional.  The classical equations of motion are Hamilton's equations \eqref{hamilton}.
Equivalently, evolution is generated by a Hamiltonian phase flow with tangent vector
\be
  V = \frac{\partial \calh}{\partial p_i}\frac{\partial}{\partial q^i}
  - \frac{\partial \calh}{\partial q^i}\frac{\partial}{\partial p_i}.
\ee

Phase space is a symplectic
manifold, which means that it naturally comes equipped with a symplectic form, which 
is a closed 2-form on $\Gamma$:
\be
  \omega = \sum_{i=1}^n \dd p_i \wedge \dd q^i \,,\qquad \dd \omega = 0\,.
\ee
The existence of the symplectic form provides us with a naturally-defined measure on phase space,
\be
  \Omega = \frac{(-1)^{n(n-1)/2}}{n!} \omega^n\,.
\ee
This is the Liouville measure, a $2n$-form on $\Gamma$.  
It corresponds to the usual way of integrating distributions over
regions of phase space,
\be
  \int f(\gamma) \Omega = \int f(q^i, p_i) d^nq d^np\,.
\ee

The Liouville measure is conserved under Hamiltonian evolution.
If we begin with a region $A \subset \Gamma$, and it evolves into a region
$A'$, Liouville's theorem states that
\be
  \int_A \Omega = \int_{A'}\Omega\,.
\ee
The infinitesimal version of this result is that the Lie derivative of $\Omega$ with
respect to the vector field $V$ vanishes,
\be
  {\cal L}_{V}\Omega = 0\,.
\ee
These results can be traced back to the fact that the original symplectic form $\omega$
is also invariant under the flow:
\be
  {\cal L}_{V}\omega = 0\,,
\ee
so any form constructed from powers of $\omega$ will be invariant.

In classical statistical mechanics, the Liouville measure can be used to assign
weights to different distributions on phase space. That is not equivalent to assigning
\emph{probabilities} to different sets of states, which requires some additional
assumption.  However, since the Liouville measure is the only naturally-defined
measure on phase space, we often assume that it is proportional to the 
probability in the absence of further information; this is essentially Laplace's ``Principle
of Indifference.''  Indeed, in statistical mechanics we typically assume that microstates
are distributed with equal probability with respect to the Liouville measure, consistent
with known macroscopic constraints.

In cosmology, we don't typically imagine choosing a random state of the universe, subject
to some constraints.  When we consider questions of fine-tuning, however, we are
comparing the real world to what we think a
randomly-chosen history of the universe would be like.
The assumption of some sort of measure
is absolutely necessary for making sense of cosmological fine-tuning arguments; otherwise
all we can say is that we live in the universe we see, and no further explanation is
needed.  (Note that this measure on the space of solutions to Einstein's equation is
conceptually distinct from a measure on observers in a multiverse, which is sometimes
used to calculate expectation values for cosmological parameters based on the
anthropic principle.)

GHS \cite{Gibbons:1986xk} showed how the Liouville measure
on phase space could be used to define a unique measure on the space of
solutions (see also \cite{Hawking:1987bi,Coule:1994gd,Gibbons:2006pa}).
In general relativity we impose the Hamiltonian constraint,
so we can consider the $(2n-1)$-dimensional constraint hypersurface of fixed Hamiltonian,
\be
  C = \Gamma/\{\calh = \calh_*\}\,.
\ee
For Robertson-Walker cosmology, the Hamiltonian precisely vanishes for either
open or closed universes, so we can take $\calh_* = 0$.  Then we consider the
space of classical trajectories within this constraint hypersurface:
\be
  M = C/V\,,
\ee
where the quotient by the evolution vector field $V$ means that two points
are equivalent if they are connected by a classical trajectory.
Note that this is well-defined, in the sense that points in $C$ always stay within $C$,
because the Hamiltonian is conserved.  

As $M$ is a submanifold of $\Gamma$, the measure is constructed by pulling 
back the symplectic form from $\Gamma$ to $M$ and raising it to the $(n-1)$th power.
GHS constructed a useful explicit form by choosing the $n$th coordinate
on phase space to be the time, $q^n=t$, so that the conjugate momentum becomes
the Hamiltonian itself, $p_n = \calh$.  The symplectic form is then
\be
  \omega = \tildeomega + \dd \calh \wedge \dd t \,,
\ee
where 
\be
  \tildeomega = \sum_{i=1}^{n-1} \dd p_i \wedge \dd q^i \,.
  \label{omegatilde}
\ee
The pullback of $\omega$ onto $C$ then has precisely the same
coordinate expression as (\ref{omegatilde}), and we will simply refer to this pullback
as $\tildeomega$ from now on.  It is automatically transverse to the Hamiltonian
flow ($\tildeomega(V)=0$), and therefore defines
a symplectic form on the space of trajectories $M$.  The associated measure is
a $(2n-2)$-form,
\be
  \ghs = \frac{(-1)^{(n-1)(n-2)/2}}{(n-1)!} \tildeomega^{n-1}\,.
  \label{ghsmeasure}
\ee
We will refer to this as the GHS measure; it is the unique measure on the space of
trajectories that is positive, independent of arbitrary choices, and respects the
appropriate symmetries \cite{Gibbons:1986xk}.

To evaluate the measure we need to define coordinates on the space of trajectories.  
We can choose a hypersurface $\Sigma$ in phase space that is transverse to the
evolution trajectories, and use the coordinates on phase space restricted
to that hypersurface.  An important property of the GHS measure is that the
integral over a region within a hypersurface is
independent of which hypersurface we chose, so long as it intersects the
same set of trajectories; if 
$S_1$ and $S_2$ are subsets, respectively, of two transverse hypersurfaces
$\Sigma_1$ and $\Sigma_2$ in $C$, with the property that the set of trajectories
passing through $S_1$ is the same as that passing through $S_2$, then
\be
  \int_{S_1} \ghs = \int_{S_2}\ghs\,.
\ee

The property that the measure on trajectories is local in phase space has a crucial implication for
studies of cosmological fine-tuning.  Imagine that we specify a certain set of trajectories
by their macroscopic properties today -- cosmological solutions that are approximately
homogeneous, isotropic, and spatially flat, suitably specified in terms of canonical 
coordinates and momenta.  It is immediately clear that the measure on this set is independent
of the behavior in very different regions of phase space, {\it e.g.} for high-density states
corresponding to early times. Therefore, no choice of early-universe Hamiltonian can make the 
current universe more or less finely tuned.  No new early-universe phenomena can change the
measure on a set of universes specified at late times, because we can always evaluate the
measure on a late-time hypersurface without reference to the behavior of the universe at
any earlier time.\footnote{On the other hand, if the effective Hamiltonian is time-dependent,
what looks like a generic state at early times can evolve into a non-generic state at later times,
as energy can be injected into the system.  This is related to the recent proposal of weak
gravity in the early universe \cite{Greene:2009tt}.}
At heart, this is a direct consequence of Liouville's theorem.

\section{Minisuperspace}
\label{minisuperspace}

In this section, we evaluate the measure on the space of solutions to Einstein's equation
in minisuperspace (Robertson-Walker) cosmology with a scalar field, applying the results to
the flatness problem and the likelihood of inflation.
We will look at two specific models:  a scalar with a canonical
kinetic term and a potential, and a scalar with a non-canonical
kinetic term chosen to mimic a perfect-fluid equation of state.

A scalar field coupled to general relativity is governed by an action
\be
  S = \int d^4x\, \sqrt{-g}\left[\frac{1}{2}R + P(X, \phi)\right],
  \label{action}
\ee
where $R$ is the curvature scalar and $P$ is the Lagrange density of the scalar field $\phi$.
We have set $\mpl^{-2} = 8\pi G = 1$ for convenience.
The scalar Lagrangian is taken to be a function of the field value and and the
kinetic scalar $X$, defined by 
\be
  X \equiv -\frac{1}{2}g^{\mu\nu}\nabla_\mu\phi \nabla_\nu\phi.
\ee
We will consider homogeneous
scalar fields $\phi(t)$ defined in a Robertson-Walker metric,
\be
  ds^2 = -N^2 \dd t^2 + a^2(t)\left[\frac{\dd r^2}{1-kr^2} + r^2 d\Omega^2\right]\,,
\ee
where the spatial curvature parameter $k$ can be normalized to $-1$, $0$, or $+1$
(so that $a(t_0)$ is not normalized to unity).  $N$ is the lapse function, which
acts as a Lagrange multiplier.  We then have
\be
  X = \frac{1}{2}N^{-2}\dot\phi^2.
\ee

\subsection{Canonical scalar field}

We start with the canonical case,
\be
  P(X,\phi) = X - V(\phi).
\ee
The Lagrangian for the combined gravity-scalar system in minisuperspace is
\be
  L = -3N^{-1}a\dot{a}^2 + 3Nak + \frac{1}{2} N^{-1}a^3 \dot\phi^2 - N a^3 V(\phi)\,.
\ee
The canonical coordinates can be taken to be the lapse function $N$, the scale factor $a$, 
and the scalar field $\phi$.  The conjugate momenta are given by
$p_i = \partial L/\partial q^i$, implying
\be
  p_N = 0\,, \quad p_a = -6N^{-1}a\dot a\,,\quad  p_\phi = N^{-1}a^3 \dot\phi\,.
  \label{momenta}
\ee
The vanishing of $p_N$ reflects the fact that the lapse function is a non-dynamical
Lagrange multiplier.
We can do a Legendre transformation to calculate the Hamiltonian, obtaining
\bea
  \calh &=&  \sum p_i\dot{q}^i - L(p_i, q^i)\\
  &=& N\left(-\frac{p_a^2}{12 a} + \frac{p_\phi^2}{2a^3} + a^3 V(\phi) - 3ak\right).
\eea
Varying with respect to $N$ gives the Hamiltonian constraint, $\calh=0$, which is
just the Friedmann equation,
\be
  H^2 = \frac{1}{3}\left(\frac{1}{2} {\dot\phi}^2 + V(\phi) - \frac{3k}{a^2}\right).
  \label{feq}
\ee

Henceforth we will set $N=1$ (consistent with the equations of motion),
leaving us with a four-dimensional phase space,
\be
  \Gamma = \{\phi, p_\phi, a, p_a\}\,.
\ee
The GHS measure on the space of trajectories is just the the Liouville measure 
subject to the constraint that $\mathcal{H}=0$,
\be 
  \ghs = (dp_a \wedge da + dp_\phi \wedge d\phi) \vert_{\mathcal{H}=0}.
  \label{minispmeasure}
\ee
Note that the measure in this example is a two-form; the full phase space is four-dimensional,
the Hamiltonian constraint surface is three dimensional, so the space of trajectories is
two-dimensional.

To express the measure in a convenient form, we use the Friedmann equation to eliminate
one of the phase-space variables.  Solving for $p_\phi$ gives us
\be
  p_\phi =
  \left[\frac{1}{6}a^{2}p_a^2 - 2a^{6}V(\phi) + 6 a^{4}k\right]^{1/2}.
\ee
We can change variables from $p_a$ to $H$ using $p_a = -6a^2H$, so that
\be
  p_\phi = \left(6a^{6}H^2- 2a^6V + 6 a^{4}k\right)^{1/2}.
\ee
Our coordinates on the constraint hypersurface $C$ are therefore $\{\phi, a, H\}$.
The basis one-forms appearing in (\ref{minispmeasure}) are 
\be
  dp_a = -12aHda -6a^2dH
\ee
and
\be
  dp_\phi = 
  \frac{6a^{4}HdH - a^{4}V' d\phi + 6a(3a^{2}H^2- a^2V
  +2k)da}{(6a^{2}H^2- 2a^2V + 6k)^{1/2}},
\ee
where $V'(\phi) = dV/d\phi$. 
Plug into the expression (\ref{minispmeasure}) for the measure, whose components become
\bea
  \ghs_{\phi H} &=& -\frac{6a^{4}}
  {\left(6a^{2}H^2- 2a^2V + 6 k\right)^{1/2}}\cr
  \ghs_{Ha} &=& -6a^2\cr
  \ghs_{a\phi} &=& 6  \frac{3a^{3}H^2- a^3V + 2ak}
  {\left(6a^{2}H^2- 2a^2V+ 6 k\right)^{1/2}}.
\eea

The measure is calculated by choosing some transverse surface $\Sigma$
in phase space, and integrating $\ghs$ over a subset of that surface.  If we choose coordinates
such that one coordinate is constant over $\Sigma$, we simply integrate the
orthogonal component of $\ghs$ with respect to the other coordinates.
One possible choice of the surface $\Sigma$ is to fix the Hubble parameter,
\be
  \Sigma : \{H = H_*\}\,.
\ee
Any consistent definition is equally legitimate; however, this choice corresponds to our
informal idea that initial conditions are set in the early universe when the Hubble parameter
is near the Planck scale.  
The measure evaluated on a surface of constant $H$ is then the integral of $\ghs_{a\phi}$,
\be
\mu = -6\int_{H=H_*}
   \frac{3a^{3}H_*^2- a^3V + 2ak}
  {\left(6a^{2}H_*^2- 2a^2V+ 6 k\right)^{1/2}}da d\phi ,
\ee
where the minus sign indicates that we have chosen an orientation that will give us
a positive final answer.
We can make this expression look more physically transparent by introducing variables
\be
  \Omega_{\dot\phi} \equiv \frac{\dot\phi^2}{6H_*^2},\quad \Omega_V\equiv \frac{V(\phi)}{3H_*^2},\quad
  \Omega_k \equiv -\frac{k}{a^2H_*^2},
\ee
so that the Friedmann equation is equivalent to
\be
  \Omega_{\dot\phi} + \Omega_V+ \Omega_k = 1.
  \label{omegasum}
\ee
The scale factor is strictly positive, so that integrating over all values of
$\Omega_k$ is equivalent to integrating over all values of $a$.  
Note that $-k/\Omega_k = 1/|\Omega_k|$.  We therefore have
\be
  da = -\frac{1}{2H_*|\Omega_k|^{3/2}}d\Omega_k,
\ee
and the measure becomes
\bea
  \mu &=&  3\sqrt{\frac{3}{2}}H_*^{-2}\int_{H=H_*} 
  \frac{1-\Omega_V - \frac{2}{3}\Omega_k}
  {|\Omega_k|^{5/2}\left(1-\Omega_V  -\Omega_k\right)^{1/2}}\,d\Omega_k d\phi \\
  &=& 3\sqrt{\frac{3}{2}}H_*^{-2}\int_{H=H_*} 
  \frac{\Omega_{\dot\phi} - \frac{1}{3}\Omega_k}
  {|\Omega_k|^{5/2}\Omega_{\dot\phi}^{1/2}}\,d\Omega_k d\phi,
  \label{flatness1}
\eea
where $\Omega_{\dot\phi}(\phi, \Omega_k)$ is defined by \eqref{omegasum}.

This integral is divergent.  One divergence clearly occurs for small values of the curvature
parameter, $\Omega_k \rightarrow 0$, as the denominator includes a factor of 
 $|\Omega_k|^{5/2}$.   The integrand also blows up at $\Omega_{\dot\phi}=0$ (or
 equivalently at $\Omega_V + \Omega_k = 1$), but the integral in that region remains
 finite.  The integral would also diverge if $\Omega_k$ or $\Omega_{\dot\phi}$ were
 allowed to become arbitrarily large, but that could be controlled by only integrating
 over a finite range for those quantities, {\it e.g.} under the theory that Planckian energy
 densities or curvatures should not be included in this classical description. 

The important divergence, therefore, is the one at $\Omega_k \rightarrow 0$, {\it i.e.}
for flat universes.  We discuss the implications of this divergence 
in Section~\ref{flatness}.

\subsection{Scalar perfect fluid}

In conventional Big Bang cosmology, we generally consider perfect-fluid sources of energy
such as matter or radiation, rather than using a single scalar field.  This situation is slightly
more difficult to analyze as a problem in phase space, as homogeneity and isotropy are only 
recovered after averaging over many individual particles.  However, we can model
a perfect fluid with an (almost) arbitrary equation of state by a scalar field with a non-canonical 
kinetic term \cite{Garriga:1999vw}. 

Consider the action (\ref{action}), where the scalar Lagrangian takes the form  
$P(X,\phi)$, where $X=-(\nabla_\mu\phi)^2/2$.  In a Robertson-Walker background,
the energy-momentum tensor takes the form of a perfect fluid,
\be
  T_\mn = (\rho+P)U_\mu U_\nu -Pg_\mn,
\ee
where the pressure is equal to the scalar Lagrange density itself (thereby accounting for the
choice of notation).  The fluid has four-velocity 
\be
  U_\mu = (2X)^{-1/2}\nabla_\mu\phi
\ee
and energy density
\be
  \rho = 2X\partial_XP - P.
\ee
We will be interested in a vanishing potential but a non-canonical kinetic term,
\be
  P(X,\phi) = \frac{2^{n-1}}{n}X^n = \frac{1}{2n}N^{-2n}\dot\phi^{2n}.
\ee
This gives a fluid with a density
\be
  \rho = \frac{2n-1}{2n}N^{-2n}\dot\phi^{2n},
\ee
corresponding to a constant equation-of-state parameter 
\be
  w = P/\rho = \frac{1}{2n-1},
\ee
as can easily be checked.  Therefore we can model the behavior of radiation ($w=1/3$) by choosing
$n=2$, and approximate matter ($w=0$) by choosing $n$ very large.

The scalar-Einstein Lagrangian in a Robertson-Walker background takes the form
\be
  L = -3N^{-1}a\dot{a}^2 + 3Nak + \frac{1}{2n}N^{-(2n-1)}a^3 \dot\phi^{2n},
\ee
and the Friedmann equation is
\be
  H^2 \equiv \left(\frac{\dot a}{a}\right)^2 = \left(\frac{2n-1}{6n}\right)\dot\phi^{2n} 
  - \frac{k}{a^2},
\ee
where we have set $N=1$.  We can duplicate the steps taken in the previous section, to 
evaluate the GHS measure in terms of coordinates $\{\phi, a, H\}$.
We end up with
\bea
  \ghs_{\phi H} &=& -\left(\frac{2n-1}{2n}\right)^{1/2n}\frac{6a^{6n/(2n-1)}}
  {\left[a^{6n/(2n-1)}3H^2 + 3 a^{2(n+1)/(2n-1)}k\right]^{1/2n}}\cr
  \ghs_{Ha} &=& -6a^2\cr
  \ghs_{a\phi} &=& 6(2n-1)^{(1-2n)/2n}
  \frac{na^{(4n+1)/(2n-1)}3H^2 + (n+1)a^{3/(2n-1)}k}
  {\left[2na^{6n/(2n-1)}3H^2 + 6n a^{2(n+1)/(2n-1)}k\right]^{1/2n}}.
\eea

To calculate the measure of a set of trajectories over a surface of constant $H=H_*$, 
we integrate $\ghs_{a\phi}$ over $a$ and $\phi$.  This yields
\be
  \mu =  -3\left(\frac{2n-1}{6n}\right)^{(1-2n)/2n}\int_{H=H_*} 
  a^2\frac{H_*^2 + \frac{(n+1)}{3n}a^{-2}k}
  {\left(H_*^2 +  a^{-2}k\right)^{1/2n}}dad\phi.
\ee
Note that the integrand has no dependence on $\phi$, since there was no potential in 
the original action.  We therefore define
\be
  x \equiv 3\left(\frac{2n-1}{6n}\right)^{(1-2n)/2n} \int d\phi,
\ee
which contributes an overall multiplicative constant to the measure.
As before, it is convenient to change variables from $a$ to $\Omega_k = -k/a^2H_*^2$.
This leaves us with
\be
  \mu = \frac{x}{2H_*^{(n+1)/n}} \int \frac{1 - \frac{(n+1)}{3n}\Omega_k}
  {|\Omega_k|^{5/2}\left(1 - \Omega_k\right)^{1/2n}}d\Omega_k.
  \label{flatness2}
\ee
This will diverge for small $\Omega_k$ for any value of $n$; all of the measure is at spatially flat universes.
This of course includes the case of radiation, $n=2$.  Therefore, the divergence we found in 
the previous subsection for flat universes does not seem to depend on the details of the
matter action.

\subsection{The flatness problem}
\label{flatness}

Let's return to the expression for the measure (\ref{flatness1}) we derived for Robertson-Walker universes
with a scalar field featuring a canonical kinetic term and a potential,
\be
  \mu \propto \int_{H=H_*} 
  \frac{1-\Omega_V - \frac{2}{3}\Omega_k}
  {|\Omega_k|^{5/2}\left(1-\Omega_V  -\Omega_k\right)^{1/2}}\,d\Omega_k d\phi. 
  \label{flatness3}
\ee
We have left out the numerical constants in front, as the overall normalization is irrelevant.
It is clear that this is non-normalizable as it stands; the integral diverges near
$\Omega_k=0$, which is certainly a physically allowed region of parameter space.
This non-normalizability is problematic if we would like to interpret the measure as
determining the relative fraction of universes with different physical properties.

We propose that the proper way of handling such a divergence is to regularize it.  That is,
we define a series of integrals that are individually finite, and which approach the original
expression as the regulator parameter $\epsilon$ is taken to zero.  We can then isolate an
appropriate power of $\epsilon$ by which we can divide the regulated expression, so that
we isolate the finite part of the result as $\epsilon$ goes to zero.

The divergence in (\ref{flatness3}) can be regulated by ``smoothing'' the factor $|\Omega_k|^{-5/2}$
in an $\epsilon$-neighborhood around $\Omega_k=0$ to get a finite integral.
Consider the function
\be
  f_\epsilon(x) =
  \begin{cases} |x|^{-5/2} & \text{if $|x| \geq \epsilon$,} \\
  \epsilon^{-5/2} &\text{if $|x| < \epsilon$.}
\end{cases}
\ee
Clearly $\lim_{\epsilon\rightarrow 0} f_\epsilon(x) = |x|^{-5/2}$, our original function.  The integral
of $f_\epsilon(x)$ over all values of $x$ is $\frac{10}{3}\epsilon^{-3/2}$.  So we obtain a
normalized integral by introducing the function
\be
  F_\epsilon(x) =
  \begin{cases} \displaystyle{\frac{3\epsilon^{3/2}}{10|x|^{5/2}}} & \text{if $|x| \geq \epsilon$,} \\
  \displaystyle{\frac{3}{10\epsilon}} &\text{if $|x| < \epsilon$,}
\end{cases}
\ee
which satisfies $\int F_\epsilon(x)dx = 1$.  We can therefore 
regularize the integral in (\ref{flatness3}) by replacing $|\Omega_k|^{-5/2}$ by 
$F_\epsilon(\Omega_k)$, and take the limit as $\epsilon \rightarrow 0$:
\be
  \mu \propto \lim_{\epsilon\rightarrow 0} \epsilon^{-3/2}\int_{H=H_*} F_\epsilon(\Omega_k)
  \frac{1-\Omega_V - \frac{2}{3}\Omega_k}
  {\left(1-\Omega_V  -\Omega_k\right)^{1/2}}\,d\Omega_k d\phi. 
  \label{flatness4}
\ee
The multiplicative factor of $\epsilon^{-3/2}$ goes to infinity in the limit, but only the finite
integral is physically relevant.  We interpret this integral as defining the normalized measure
on the space of cosmological spacetimes.

However, it is clear that the limit of $F_\epsilon(x)$ is simply a delta function,
\be
  \lim_{\epsilon\rightarrow 0}F_\epsilon(x) = \delta(x),
\ee
in the sense that the integral over a test function $\psi(x)$ gives
\be
  \lim_{\epsilon\rightarrow 0} \int_{-\infty}^{\infty}F_\epsilon(x)\psi(x)\, dx = \psi(0). 
\ee
Consequently, the measure is entirely concentrated on exactly flat universes; universes with
nonvanishing spatial curvature are a set of measure zero.  The
integrated measure (\ref{flatness4}) is equivalent to
\be
  \mu \propto \int_{H=H_*} 
  \sqrt{1-\Omega_V}\, d\phi,
  \label{flatness5}
\ee
with $\Omega_k$ fixed to be $0$.

Therefore, our interpretation is clear:  almost all universes are spatially flat.  In terms of the
measure defined by the classical theory itself, a ``randomly chosen'' cosmology will be
flat with probability one.  The flatness problem, as conventionally understood, does not 
exist; it is an artifact of informally assuming a flat measure on the space of initial cosmological 
parameters.  Of course, any particular specific theory of initial conditions might actually
have a flatness problem, if it predicts spatially-curved universes with high probability; but
that problem is not intrinsic to the standard Big Bang model by itself.

Classical general relativity is not a complete theory of gravity, and
our notions of what constitutes a ``natural'' set of initial conditions are inevitably informed
by our guesses as to how it will ultimately be completed by quantum gravity.  At the level
of the classical equations of motion, initial data for a solution may be specified at any time;
Hamilton's equations then define a unique solution for the complete past and future.
However, we generally impose a cutoff on the validity of a classical solution when some
quantity -- the energy density, Hubble parameter, or spatial curvature -- reaches the
Planck scale.  It therefore makes sense to us to imagine that some unknown physical process
sets the initial conditions near the Planck regime.  In Robertson-Walker cosmology, we
might imagine that the space of allowed initial conditions consists of all values of the
phase-space variables such that the energy density and curvatures are all sub-Planckian;
in terms of the density parameters $\Omega_i$, this corresponds to $|\Omega_{i,{\rm pl}}|<1$,
where the subscript ``pl'' denotes that the quantity is evaluated when $H\sim \mpl$.

\begin{figure}[t]
\begin{center}
\includegraphics[width=0.8\textwidth]{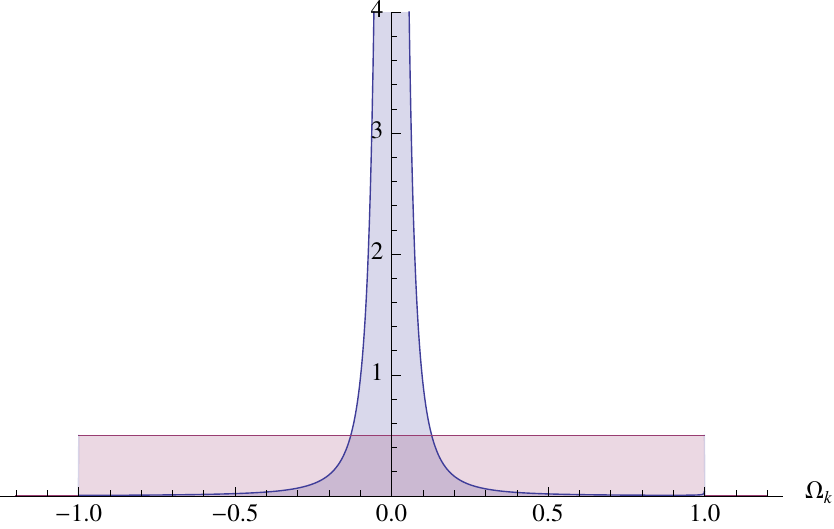}
\end{center}
\caption{Two measures as a function of the curvature parameter $\Omega_k$. 
The GHS measure is highly peaked near the origin, indicating a
divergence at spatially flat universes. (We've drawn the unnormalized measure; a normalized
version would simply be a $\delta$-function.)  This is in stark contrast with the flat 
distribution generally assumed in the discussion of the flatness problem, which we've
plotted for $\Omega_k$ between $\pm 1$.}
\label{measureplot}
\end{figure}

What this means in practice is that we tend to assign equal probability -- a flat prior -- to all the
allowed $\Omega_{i,{\rm pl}}$'s when contemplating cosmological initial conditions.
As a matter of principle, it is necessary to invoke some kind of prior in order to 
sensibly discuss fine-tuning problems; a quantity is finely-tuned if it is drawn from 
a small (as defined by some measure) region of parameter space.
The lesson of the GHS measure is that a flat prior on $\Omega_{i,{\rm pl}}$ ignores the structure of the
classical theory itself, which comes equipped with a unique well-defined measure.
In Figure~\ref{measureplot} we plot two different measures on the value of 
$\Omega_k$ at the Planck scale; the informal flat prior assumed in typical discussions of
the flatness problem, and the GHS measure (evaluated at $\Omega_V=0$ for convenience).
We see that using the measure defined by the classical equations of motion leads to a 
dramatic difference in the probability density.

Note that the model of a canonical scalar field with a potential will allow for the possibility
of inflation if the potential is chosen appropriately; however, the divergence at flat universes
is \emph{not} because inflation is secretly occurring.  For one thing, the divergence
appears for any choice of potential, and also in the perfect-fluid model where there is
no potential at all.  For another, we could always choose to evaluate the measure at late
times -- {\it i.e.}, we could pick the fixed Hubble parameter $H_*$ to be very small.
The measure on trajectories is independent of this choice, 
so the divergence for flat universes cannot depend on whether inflation occurs.

This divergence was noted in
the original GHS paper \cite{Gibbons:1986xk}, where it was attributed to ``universes with
very large scale factors'' due to a different choice of variables.  This is not the most physically
transparent characterization,  as any open universe will eventually have a large scale factor. 
It is also discussed by Gibbons and Turok \cite{Gibbons:2006pa}, who correctly attribute it
to nearly-flat universes.  However, they advocate discarding all such universes as
physically indistinguishable, and concentrating on the non-flat universes.  To us, this seems
to be throwing away almost all the solutions, and keeping a set of measure zero.  It is true
that universes with almost identical values of the curvature parameter will be physically
indistinguishable, but that doesn't affect the fact that almost all universes have this property.
In Hawking and Page \cite{Hawking:1987bi} and Coule \cite{Coule:1994gd} the divergence
is directly attributed to flat universes, but they do not seem to argue that the flatness problem
is therefore an illusion.

The real world is not precisely Robertson-Walker, so in some sense the flatness problem
is not rigorously defined; a super-Hubble-radius perturbation could lead to a deviation from
$\Omega=1$ in our observed universe, even if the background cosmology were spatially flat.
Nevertheless, the unanticipated structure of the canonical measure in minisuperspace serves as a cautionary
example for the importance of considering the space of initial conditions in a
rigorous way.  More directly, it raises an obvious question:  if the canonical measure is
concentrated on spatially flat universes, might it also be concentrated on smooth universes,
thereby calling into question the status of the horizon problem as well as the flatness problem?
(We will find that it is not.)

\subsection{Likelihood of inflation}

A common use of the canonical measure has been to calculate the likelihood of inflation
\cite{Gibbons:1986xk,Hawking:1987bi,Coule:1994gd}.  Most recently, Gibbons and Turok
\cite{Gibbons:2006pa}  have argued that the fraction of universes that inflate is extremely small.
However, they threw away all but a set of measure zero of trajectories, on the grounds that they all
had negligibly small spatial curvature and therefore physically indistinguishable.  Inflation, of
course, tends to make the universe spatially flat, so this procedure is potentially unfair to the
likelihood of inflation.  We therefore re-examine this question, following the philosophy suggested
by the above analysis, which implies that almost all universes are spatially flat.  We
choose to look only at flat universes, and calculate the fraction that experience more than sixty 
$e$-folds of inflation.  We will look at two choices of potential:  a massive scalar,
and a pseudo-Goldstone boson.  (We will argue in the next section that these results are
physically irrelevant, as perturbations play a crucial role.)

We start with a massive scalar field with $m_{\phi}=3 \times 10^{-3}\mpl$, which yields an 
amplitude of perturbations that agrees with observations. 
We choose to evaluate the measure on the hypersurface $H=1/\sqrt{3}$, so that the Friedmann 
equation becomes $1=\frac{1}{2}\dot\phi^2+\frac{1}{2}m^2\phi^2$. 
After replacing the divergence at zero curvature in \eqref{flatness2} by a delta function, the normalized measure becomes
\begin{equation}\label{infmeasure}
\mu = \frac{\sqrt{2}m}{\pi} \int_{H=1/\sqrt{3}} \sqrt{1-\frac{1}{2}m^2\phi^2} \,d\phi.
\end{equation}
(Recall that we have set $\mpl = 1/\sqrt{8\pi G}=1$.)  The range of integration is
$|\phi| \leq \sqrt{2/m_\phi^2}$ (or $\rho_{\dot\phi} \leq 1$), corresponding to $V \leq 1$. 

We used the Euler method with a time step $\Delta t = 10^{-3}$ to numerically follow the evolution of the 
scale factor and the scalar field.  We
find that the universe undergoes more than sixty $e$-folds of inflation for all initial values 
of $\phi$ except for the range $-24$ to $6$ if $\dot\phi > 0$ (for $\dot\phi < 0$, the range would be $-6$ 
to $24$ due to the symmetry of the potential). For simplicity, we disregard in our calculation the expansion 
that occurs after the first period of slow-roll inflation. 
(We verified numerically that subsequent periods of slow-roll expansion are relatively brief and 
lead to very little further expansion.) Excluding the region $-24 \leq \phi \leq 6$, the measure \eqref{infmeasure} integrates to 0.99996. It seems highly likely to have more than sixty $e$-folds of inflation by this standard. 

As another example we consider inflation driven by
a pseudo-Goldstone boson \cite{Freese:1990rb} with potential 
\be
  V(\phi) = \Lambda^4(1+\cos(\phi/f)).
\ee 
In our calculation, we use $f=\sqrt{8\pi}$ and $\Lambda = 10^{-3}$, so that the model is 
consistent with WMAP3 data \cite{Freese:2008if}. We evaluate the measure on the 
hypersurface $H=H_*=\sqrt{4/3}\Lambda^2$, so that $3H_*^2 = 2V_{max} = 4\Lambda^4$. In this case, the normalized measure becomes
\begin{equation}
  \mu = \frac{1}{8\sqrt{\pi}E[-1]}
  \int_{H_*=\sqrt{4/3}\Lambda^2}\sqrt{1-\frac{1}{4}\left(1+\cos{\frac{\phi}{\sqrt{8\pi}}}\right)} d\phi,
\end{equation}
where $E[m]$ is the elliptic integral $\int_0^{2\pi}\sqrt{1-m \sin^2{t}}dt$. Numerically we find that the universe 
expands by more than $60$ efolds  for $-4.0<\phi<2.4$ if $\dot\phi>0$ at $H=H_*=\sqrt{4/3}\Lambda^2$. (The
evenness of the potential allows us to consider only this branch of solutions.) Evaluating the measure gives a
probability of $0.171$. Notice that this is not too different from the calculation in \cite{Freese:1990rb}, which gives
$0.2$ by assuming that $\phi$ is randomly distributed between $0$ and $\sqrt{8\pi}$. We also note that the 
probability is rather sensitive to the value of $f$; numerical evidence suggest that it increases with $f$ (a flatter 
potential). 

Both of these examples lead to the conclusion that inflation has a very reasonable chance of occurring.
Indeed, it is sometimes claimed that inflation is an ``attractor'' (see {\it e.g.} \cite{Trodden:2004st}),
but that is a misleading abuse of nomenclature.  It is a basic feature of dynamical systems theory that
there are no attractors in true Hamiltonian mechanics; Liouville's theorem implies that the total
volume of a region of phase space remains constant under time evolution.  Attractors, in the
rigorous sense of the word, only occur for
systems with dissipation.  Inflation appears to be an attractor only because it is often convenient to 
portray ``phase portraits'' in terms of the inflaton $\phi$ and its time derivative, $\dot\phi$.
But $\dot\phi$ is not the momentum conjugate to $\phi$; as seen in \eqref{momenta}, with
the lapse function set to $N=1$, it is $p_\phi = a^3\dot\phi$.  Trajectories drawn on a $(\phi, \dot\phi)$
plot tend to approach a fixed point, but only because the scale factor $a$ is dramatically increasing,
not because of any true attractor behavior.

These calculations of the likelihood of inflation are of dubious physical relevance.
Examining a single scalar field in minisuperspace is an extremely unrealistic scenario.
At a very simple level, if there are other massless fields in the problem, any of them may share some of
the energy density, reducing the probability that the inflaton potential dominates.  More importantly,
the role of perturbations is crucial.  The real reason why inflation is unlikely from the point of view of the
canonical measure is not because it is unlikely in minisuperspace, but because perturbations can easily
be sufficiently large to prevent inflation from ever occurring.  We examine this issue in detail in the next section.

\section{Perturbations}
\label{perturbations}

The horizon problem is usually formulated 
in terms of the absence of causal contact between widely-separated points in the early universe.
Operationally, however, it comes down to the fact that the universe is smooth over large
scales.  We can investigate the measure associated with such universes by looking at
perturbed Robertson-Walker cosmologies.  While the set of all perturbations defines a
large-dimensional phase space, in linear perturbation theory 
we can keep things simple by looking at a single mode at a time.
We will find that, in contrast with the surprising result of the last section, the measure on 
perturbations is just what we would expect -- there is no divergence at nearly-smooth
universes.  However, this implies that only an imperceptibly small fraction of spacetimes were 
sufficiently smooth at early times to allow for inflation to occur.

To calculate the measure for scalar perturbations, we need to first compute the corresponding action. 
We are interested in universes dominated by hydrodynamical matter such as dust or radiation.
For linear scalar perturbations, the coupled gravity-matter system can be described by a single
independent degree of freedom, as discussed by Mukhanov, Feldman and Brandenberger 
\cite{Mukhanov:1990me}; we will follow closely the discussion in \cite{Mukhanovtext}.
After obtaining the action, we can isolate the dynamical variables and construct 
the symplectic two-form on phase space, which can then be used to compute the measure on the 
set of solutions to Einstein's equations.  A slight subtlety arises because the corresponding
Hamiltonian is time-dependent, but this is easily dealt with.

\subsection{Description of perturbations}

In this section it will be convenient to switch to conformal time,
\be
  \eta = \int a^{-1}dt.
\ee
Derivatives with respect to $\eta$ are denoted by the superscript $'$, 
and $\htilde \equiv a'/a$ is related to the Hubble parameter $H=\dot{a}/a$ by
$\htilde = aH$.  The Friedmann equations become
\bea
  \htilde^2 &=& \frac{8\pi G}{3}a^2\bar\rho - k,\\
  \htilde' &=& -\frac{4\pi G}{3}a^2(\bar\rho+3\bar{p}),
\eea
where $\bar\rho$ and $\bar{p}$ are the background density and pressure.
In a flat universe with only matter and radiation, in the radiation-dominated era we have
\be
 \eta({\rm RD}) = \frac{a}{H_0\sqrt{a_\eq}} ,\qquad \htilde({\rm RD}) =\eta^{-1},
\ee
where $a_\eq$ is the scale factor at matter-radiation equality, and now we set the current scale
factor to unity, $a_0=1$.
Numerically, the conformal time in the radiation-dominated era is approximately
\be
  \eta(T) \approx \frac{5\times 10^{30}}{T({\rm eV})} \,{\rm eV}^{-1}.
\ee

The metric for a flat RW universe in conformal time with scalar perturbations is
\be \label{metricscalar}
ds^2 = a^2(\eta)\left[-(1+2\Phi)d\eta^2 + 2B_{,i}d\eta dx^2 + ((1-2\Psi) \delta_{ij}+2E_{,ij})dx^i dx^j\right],
\ee
where $\Phi$, $\Psi$, $E$, and $B$ are scalar functions characterizing metric perturbations, and 
commas denote partial derivatives. It is useful to define the gauge-invariant Newtonian potential,
\be
  \Phi = \phi - \frac{1}{a}\left[a(B-E)'\right]',
\ee
and the gauge-invariant energy-density perturbation,
\be
  \drt = \delta\rho - \bar{\rho}'(B-E').
\ee
For scalar perturbations in the absence of anisotropic stress, these are related by
\be
  \drt = \frac{1}{4\pi G a^2}\left[\nabla^2\Phi - 3\htilde(\Phi' + \htilde\Phi)\right].
  \label{drhoeq}
\ee

For adiabatic perturbations ($\delta S=0$), the potential obeys an autonomous equation,
\be
  \Phi'' + 3(1+c_s^2)\htilde\Phi' -c_s^2\nabla^2\Phi + [2\htilde' + (1+3c_s^2)\htilde^2]\Phi = 0,
  \label{phieq}
\ee
where $c_s^2=\partial p/\partial\rho$ is the speed of sound squared in the fluid.
This equation simplifies if we introduce the rescaled perturbation variable
\be
  u \equiv \frac{\Phi}{\sqrt{\bar\rho + \bar{p}}},
  \label{uphi}
\ee
and the time-dependent parameter
\be
  \theta = \exp\left[\frac{3}{2}\int (1+c_s^2)\htilde d\eta\right]\Phi
  =\frac{1}{a}\left[\frac{2}{3}\left(1 - \frac{\htilde'}{\htilde^2}\right)\right]^{-1/2}.
\ee
In terms of these \eqref{phieq} becomes
\be
  u'' -c_s^2\nabla^2u - \frac{\theta''}{\theta} u = 0.
  \label{ueq}
\ee

The variable $u$ is a single degree of freedom that encodes both 
the gravitational potential [through \eqref{uphi}] and the density perturbation [through \eqref{drhoeq}].
The equation of motion \eqref{ueq} corresponds to an action
\be
  S_u = \frac{1}{2}\int d^4x \left(u'^2 - c_s^2 u_{,i}u_{,i} + \frac{\theta''}{\theta}u^2 \right).
\ee
Defining the conjugate momentum $p_u = \partial{\mathcal L}/\partial u' = u'$, we can describe the
dynamics in terms of a Hamiltonian density for an individual mode with wavenumber $k$, 
\begin{equation}
  \mathcal{H} = \frac{1}{2}p_u^2 + \frac{1}{2}\left(c_s^2k^2  - \frac{\theta''}{\theta}\right)u^2.
  \label{pertham}
\end{equation}
This is simply the Hamiltonian for a single degree of freedom with a time-dependent
effective mass $m^2= c_s^2k^2  - \theta''/\theta$.

\subsection{Computation of the measure}

Given the Hamiltonian \eqref{pertham}, we can straightforwardly compute the invariant measure on phase space. One caveat is that now the Hamiltonian is time-dependent, because the effective mass evolves.
The carrier manifold of the Hamiltonian therefore has an odd number of dimensions. We can retain the 
symplecticity of a time-dependent Hamiltonian system (which requires an even number of dimensions) 
by promoting time to be an addition canonical coordinate, $q^{n+1}=t$. The conjugate momentum is 
minus the Hamiltonian, $p_{n+1}=-\mathcal{H}$. We can then define an extended Hamiltonian by
\be
  \mathcal{H}_+ = \mathcal{H}(p,q,t)+p_{n+1}.
\ee  
This is formally time-independent, and recovers the original Hamiltonian equations via
\be
  \dot{q}_i=\frac{\partial \mathcal{H}_+}{\partial p^i}\,, \quad \dot{p}^i=-\frac{\partial \mathcal{H}_+}{\partial q^i},
\ee
along with two additional trivial equations $\dot{t}=1$ and $\dot{\mathcal{H}}=\partial \mathcal{H}/\partial t$. 

With $t$ promoted to a coordinate, the time-dependent Hamiltonian system also comes equipped naturally with a closed symplectic two-form, now with an additional term:
\begin{equation}\label{extsymform}
\omega = \sum_{i=1}^{n} dp_i \wedge dq^i - d\mathcal{H} \wedge dt. 
\end{equation} 
The invariance of the form of Hamilton's equations ensures that the Lie derivative of $\omega$ with respect to the vector field generated by $\mathcal{H}_+$ vanishes. The top exterior power of $\omega$ is then guaranteed to be conserved under the extended Hamiltonian flow, and can thus play the role of the Liouville measure for the augmented system. The GHS measure can then be obtained by pulling back the Liouville measure onto a hypersurface intersecting the trajectories and satisfying the constraint $\mathcal{H}_+=0$. 

In our case, the original system, with coordinate $u$ and conjugate momentum $p_u$, is augmented to one with two coordinates $u$ and $\eta$ and their conjugate momenta $p_u$ and $-\mathcal{H}$. The extended Hamiltonian, 
\be
  \mathcal{H}_+ = \frac{1}{2}p_u^2+\frac{1}{2}\left(c_s^2k^2-\frac{\theta''}{\theta}\right)u^2-\mathcal{H},
\ee 
is time-independent and set to zero by the equations of motion. Its conservation is analogous to the Friedmann equation constraint in the analysis of the flatness problem.  
Using \eqref{extsymform}, the GHS measure $\Theta$ for the perturbations is the two-form
\begin{eqnarray} \nonumber
  \Theta &=& dp_u \wedge du - (d\mathcal{H} \wedge d\eta)|_{\mathcal{H}_+=0} \\ \nonumber
  &=& dp_u \wedge du - \frac{1}{2}d\left[p_u^2 
     + \left(c_s^2 k^2 - \frac{\theta''}{\theta}\right)u^2\right] \wedge d\eta \\ \label{measure}
  &=& dp_u \wedge du - p_u(dp_u \wedge d\eta) - u\left(c_s^2 k^2 - \frac{\theta''}{\theta}\right)du \wedge d\eta\,.
\end{eqnarray}

One convenient hypersurface in which we can evaluate the flux of trajectories is $\eta=
\eta_*=\mbox{constant}$.  (This is equivalent to a surface of $H=\mbox{constant}$ or
$a=\mbox{constant}$, although those are not coordinates in the phase space of the perturbation.)
As $\eta$ is always positive in a matter- and radiation-dominated universe, this surface intersects all 
trajectories exactly once.  We then have
\bea
  \mu &=& \int_{\eta=\eta_*} \Theta_{p_uu} dudp_u \nonumber \\
  &=& \int_{\eta=\eta_*} dudp_u.
  \label{pertmeasure}
\eea
The flux of trajectories crossing this surface is unity, implying that all values for 
$u$ and $p_u$ are equally likely. There is nothing in the measure that would explain the small 
observed values of perturbations at early times.  
Hence, the observed homogeneity of our universe does imply considerable fine-tuning; unlike the
flatness problem, the horizon problem is real.

\subsection{Likelihood of inflation}

We can use the canonical measure on perturbations to estimate the likelihood of 
inflation.   Our strategy will be to consider universes dominated by matter and radiation -- {\it i.e.},
the supposed post-inflationary era in the universe's history -- and ask what fraction of them could have 
begun with inflation.  This is somewhat contrary to the conventional approach, which might start
with an assumed early state of the universe and ask whether inflation will begin; but it is fully
consistent with the philosophy of unitary and autonomous evolution, and takes advantage of
the feature of the canonical measure that it can be evaluated at any time.

If inflation does occur, perturbations will be very small when it ends.  Indeed, perturbations must
be sub-dominant if inflation is to begin in the first place \cite{Vachaspati:1998dy}, and by the end
of inflation only small quantum fluctuations in the energy density remain.  It is therefore a 
necessary (although not sufficient) condition for inflation to occur that perturbations be small
at early times.  For convenience, we will take inflation to end near the GUT scale,
$T_{\rm G} = 10^{16}$~GeV ($\eta_{\rm G} \approx 6\times 10^{5}\,{\rm eV}^{-1}$), 
although this choice is not crucial.

We therefore want to calculate what fraction of perturbed Robertson-Walker universes are
relatively smooth near the GUT scale.  We take ``smooth" to mean that both
the density contrast $\delta = \drt/\bar\rho$ and the Newtonian potential $\Phi$ are less than one.
Because the phase space is unbounded and the measure \eqref{pertmeasure} is flat,
it is necessary to cut off the space of perturbations in some way.   We might define ``realistic''
cosmologies as those that match the homogeneity of our observed universe at the redshift of
recombination $z\sim 1200$, when CMB temperature anisotropies are observed.
Our expressions will be much less cumbersome, however, if we demand smoothness at
matter-radiation equality, $z_\eq \sim 3000$ ($\eta_{\eq} \approx  10^{31}\,{\rm eV}^{-1}$), 
within an order of magnitude of recombination.
Since the observed temperature anisotropies are of order one part in $10^5$,
we therefore define a realistic universe as one with $\delta_\eq \leq 10^{-5}$ and
$\Phi_\eq \leq 10^{-5}$.

There is a long-distance cutoff on the modes we consider given by the size of our comoving observable
universe, extrapolated back to matter-radiation equality.
The size $L_0$ of our observable universe today is a few times $H_0^{-1} = 10^{33}~{\rm eV}^{-1}$, 
and the size of our comoving patch at equality is $a_\eq = 1/3000$ times that, so
\be
  L_\eq \approx 10^{30}\,{\rm eV}^{-1}.
\ee
We will also impose a short-distance cutoff at the Hubble radius at equality,
\be
  H_\eq^{-1} \approx \mpl\left(\frac{a_\eq^2}{T_0^2}\right) \approx 10^{28}\,{\rm eV}^{-1}.
\ee
The total number of modes we consider is therefore
\be
  n = \left(\frac{L_\eq}{H_\eq^{-1}}\right)^3 \approx 10^6.
\ee
Our short-distance cutoff is chosen primarily for convenience; there is a natural 
ultraviolet cutoff set by the scale below which the hydrodynamical approximation becomes invalid,
but that is much shorter than $H_\eq^{-1}$.  It is clear
that we are neglecting a large number of modes that could plausibly have large amplitudes
at early times; our result will therefore represent a generous overestimate of the fraction of 
inflationary spacetimes.  The final numerical answer will be small enough that this shortcut won't matter.

With this setup in place, we would like to compare the measure on trajectories that are smooth
near the GUT scale to the measure on those that are realistic near matter-radiation equality.
We therefore only need to consider 
a single kind of evolution -- long-wavelength modes (super-Hubble-radius, $c_sk\eta < 1$) 
in a radiation-dominated universe.  In that case the general solution to our 
evolution equation \eqref{ueq} is
\be
  u = c_1\theta + c_2\theta \int_{\eta_0}\theta^{-2} \, d\eta,
  \label{solution}
\ee
where $c_1$ and $c_2$ are constants.  During radiation domination we have
\be
  \theta =\frac{\sqrt{3}}{2\sqrt{a_\eq}H_0}\eta^{-1}.
\ee
The solution for $u$ is therefore
\be
  u = \alpha \eta^{-1} + \beta \eta^{2},
  \label{ualphabeta}
\ee
where $\alpha$ and $\beta$ are constants.  The conjugate momentum is
\be
  p_u = -\alpha \eta^{-2} + 2\beta \eta.
  \label{pualphabeta}
\ee

The potential is related to $u$ by \eqref{uphi}.  In the radiation era we have 
\be
  (\bar\rho + \bar{p})^{1/2} = \gamma\eta^{-2},
\ee
where we have defined
\be
  \gamma = \frac{2\mpl}{\sqrt{a_\eq}H_0}.
\ee
Our general solution is therefore
\be
  \Phi = \gamma(\alpha\eta^{-3} + \beta).
  \label{phialphabeta}
\ee
Finally we turn to the density perturbation, which is given by \eqref{drhoeq}.
The $\nabla^2\Phi=-k^2\Phi$ term is negligible for long wavelengths, so we're left with
\be
  \drt = \frac{12\mpl^3}{a_\eq^{3/2}H_0^3}(2\alpha\eta^{-7} - \beta\eta^{-4}),
\ee
which in turn implies
\be
  \delta \equiv \frac{\drt}{\bar{\rho}} = 2\gamma(2\alpha\eta^{-3} - \beta).
    \label{rhoalphabeta}
\ee

To calculate the measure, it is convenient to use $\alpha$ and $\beta$ as the independent variables
that specify a mode.  The measure is simply
\be
  \mu = \int du dp_u = 3\int d\alpha d\beta.
\ee
This comes from taking the derivative of \eqref{ualphabeta} and \eqref{pualphabeta}, treating
$\alpha$ and $\beta$ as the independent variables, and computing $du \wedge dp_u$.
No $\eta$-dependent factors appear when we write the measure in terms of $\alpha$ and $\beta$.  
We can also express it in terms of the density contrast and Newtonian potential,
\be
  d\Phi d\delta = 6\gamma^2\eta^{-3}d\alpha d\beta.
\ee
Therefore, a region in the $\Phi$-$\delta$ plane at time $\eta_{\rm G}$ has a measure that is larger
than the same coordinate region at time $\eta_\eq$ by a factor of
\be
  \left(\frac{\eta_\eq}{\eta_{\rm G}}\right)^3 \approx 10^{76} .
\ee

The coordinate area of our initial region at the GUT scale is $\Delta\Phi_{\rm G} \Delta\delta_{\rm G} \approx 1$, 
while the coordinate area of our region at equality is $\Delta\Phi_{\eq} \Delta\delta_{\eq}
\approx (10^{-5})^2 = 10^{-10}$.  For each mode, we therefore have
\be
  \frac{\mu(\mathrm{inflationary})}{\mu(\mathrm{realistic})} =
  \left(\frac{\eta_{\rm G}}{\eta_\eq}\right)^3
  \frac{\Delta\Phi_{\rm G} \Delta\delta_{\rm G}}{\Delta\Phi_{\eq} \Delta\delta_{\eq}} \approx 10^{-66}.
\ee
This is saying that, for a given wave vector, only $10^{-66}$ of the allowed amplitudes that are
realistic at matter-radiation equality are small at the GUT scale.  
To allow for inflation, we require that modes of every fixed comoving wavelength and direction
be less than unity at the GUT scale;
the fraction of realistic cosmologies that are eligible for inflation is therefore
\be
  P({\rm inflation}) \approx
  (10^{-66})^n \approx 10^{-6.6\times 10^7}.
\ee

This is a small number, indicating that a negligible fraction of universes that are
realistic at late times experienced a period of inflation at very early times.  
We derived this particular value by assuming the universe was realistic at matter-radiation equality,
but similarly tiny fractions would apply had we started with any other time in the late universe.
We also looked at only a fraction of possible modes, so a more careful estimate would
yield a much smaller number.  Indeed, using entropy as a proxy for the number of states yields
estimates of order $10^{-10^{122}}$ \cite{penrose}.  Clearly, the precise numerical answer
is not of the essence; the conclusion is that inflationary trajectories are a negligible fraction of
all possible evolutions of the universe.

A crucial feature
of this analysis is that we allowed for the possibility of \emph{decaying} cosmological perturbations;
if all we know about the perturbations is that they are small at matter-radiation equality, the generic
case is that many have been decaying since earlier times.  Such decaying modes are often neglected
in cosmology, but for our purposes that would be begging the question.  A successful theory of
cosmological initial conditions will account for the absence of such modes, not presume it.

\section{Discussion}

We have investigated the issue of cosmological fine-tuning under the assumption that our 
observable universe evolves unitarily through time.  
Using the invariant measure on cosmological solutions to Einstein's equation, we find that the
flatness problem is an illusion; in the context of purely Robertson-Walker cosmologies, the measure 
diverges on flat universes.  In the case of deviations from homogeneity, however, we recover something
closer to the conventional result; in appropriate variables, the measure on the phase space of
any particular mode of perturbation is flat, so that a generic universe would be expected to be 
highly inhomogeneous.

Inflation by itself cannot solve the horizon problem, in the sense of making the smooth early universe
a natural outcome of a wide variety of initial conditions.  The assumptions of unitarity and autonomy 
applied to our comoving patch imply that
any set of states at late times necessarily corresponds to an equal number of states at
early times.   Different choices for the Hamiltonian relevant in the early universe
cannot serve to focus or spread the trajectories, which would violate 
Liouville's theorem; they can only deflect the trajectories in some overall
way.  Therefore, whether or not a theory allows for inflation has no
impact on the total fraction of initial conditions that lead to a universe that looks like
ours at late times.

This basic argument has been appreciated for some time; indeed, its essential features
were outlined by Penrose \cite{penrose} even before inflation was invented.  
Nevertheless, it has failed to make an important impact on most discussions of
inflationary cosmology.  Attitudes toward this line of inquiry fall roughly into three
camps:  a small camp who believe that the implications of Liouville's theorem represent
a significant challenge to inflation's purported ability to address fine-tuning
problems \cite{Unruh:1996sf,Hollands:2002yb,Hollands:2002xi,Carroll:2004pn,Gibbons:2006pa};
an even smaller camp who explicitly argue that the allowed space of initial conditions is
much smaller than the space of later conditions, in apparent conflict with 
the principles of unitary
evolution \cite{Kofman:2002cj,Linde:2007fr}; and a very large camp who choose to 
ignore the issue or keep their opinions to themselves.  

But this issue is crucial to understanding the role of inflation (or any alternative mechanism)
in accounting for the apparent fine-tuning of our universe.
The part of the universe we observe consists of a 
certain set of degrees of freedom, arranged in a certain way --  a few hundred billion galaxies,
distributed approximately uniformly through an expanding space -- and apparently evolving
from a very finely-tuned smooth Big-Bang-like beginning.  Understanding why things are this
way could have crucial consequences for our view of other features of the universe, much as
the inflationary scenario revolutionized our ideas about the origin of cosmological perturbations.

There seem to be two possible ways we might hope to account for the apparent fine-tuning
of the history of the observable universe:
\begin{enumerate}
\item The present configuration of the universe only occurs once.  In this case, the evolution from
the Big Bang to today is highly non-generic, and the question becomes why this evolution, rather
than some other one.  The answer might be found in properties of the wave function of the
universe ({\it e.g.} \cite{Hartle:2008ng}).
\item Degrees of freedom arrange themselves in configurations like the observable universe many
times in the history of a much larger multiverse.  In this case, there is still hope that the overall
evolution may be generic, if it can be shown that configurations like ours most often occur in the
aftermath of a smooth Big Bang.  The apparent restrictions of Liouville's theorem may be
circumvented by imagining that the degrees of freedom of our current universe do not 
describe a closed system for all time, but interact strongly with other degrees of freedom at some times
({\it e.g.} by arising as baby universes \cite{Carroll:2004pn}).
\end{enumerate}

In either case, inflation could play an important role as part of a more comprehensive picture.
While inflation does not make universes like ours more
numerous in the space of all possible universes, it might provide a more 
reasonable target for a true theory of initial conditions, from quantum cosmology or
elsewhere.  (This is a possible reading of \cite{Kofman:2002cj,Linde:2007fr}, although
those authors seem to exclude non-smooth initial conditions {\it a priori}, rather than 
relying on some well-defined theory of initial conditions.)

As we have shown in this paper, most universes that are smooth at matter-radiation equality were
wildly inhomogeneous at very early times.  But the converse is not true; most universes that were
wildly inhomogeneous at early times simply stay that way.  The process of smoothing out represents
a violation of the Second Law of Thermodynamics, as entropy decreases along the 
way.  Even though the vast majority of trajectories that are smooth at matter-radiation
equality were inhomogeneous at early times, it seems intuitively unlikely that the real universe 
behaves this way; much more plausible is the conventional supposition that the universe was smoother 
(and entropy was lower) all the way back to the Big Bang.

One way of expressing why this seems more natural to us is that the corresponding initial
states are very simple to characterize:  they are smooth within an appropriate comoving volume.
In contrast, the much more numerous histories that begin inhomogeneously and proceed to smooth
out are impossible to characterize in terms of macroscopically observable quantities at early times;
the fact that they will ultimately smooth out is hidden in extremely
subtle correlations between a multitude of degrees of freedom.\footnote{The situation resembles 
the time-reversal of a glass of water with an ice cube that melts
over the course of an hour.  At the end of the melting process, if we reverse the momentum
of every molecule in the glass, we will describe an initial condition that evolves into an
ice cube.  But there's no way of knowing that, just from the macroscopically available
information; the surprising future evolution is hidden in subtle correlations between
different molecules.}  It seems much easier to imagine
that an ultimate theory of initial conditions will produce states that are simple to describe
rather than ones that feature an enormous number of mysterious and inaccessible
correlations.  It may be true that a randomly-chosen universe like ours would have begun
in a wildly inhomogeneous state; but it's clear that the history of our observable universe 
is not a randomly-chosen evolution of the corresponding degrees of freedom.  

Given that we need some theory of initial conditions to explain why our universe
was not chosen at random, the question becomes whether inflation provides
any help to this unknown theory.  There are two ways in which it does.  First, inflation allows the
initial patch of spacetime with a Planck-scale Hubble parameter to be physically small,
while conventional cosmology does not.  If we extrapoloate 
a matter- and radiation-dominated universe from today backwards in time,
a comoving patch of size $H_0^{-1}$ today corresponds to a physical size
$\sim 10^{-26}H_0 \sim 10^{34}\lpl \sim 1$~cm when $H=\mpl$.  In contrast, with inflation,
the same patch needs to be no larger than $\lpl$ when $H=\mpl$,
as emphasized by Kofman, Linde, and Mukhanov \cite{Kofman:2002cj,Linde:2007fr}.
If our purported theory of initial conditions, whether quantum cosmology or baby-universe
nucleation or some other scheme, has an easier time making small patches of space
than large ones, inflation would be an enormous help.

The other advantage is in the degree of smoothness required.  Without inflation,
a perfect-fluid universe with Planckian Hubble parameter
would have to be extremely homogeneous to be compatible with the current universe,
while an analogous inflationary patch could accommodate any amount of 
sub-Planckian perturbations.  While the actual number of trajectories may be 
smaller in the case of inflation, there is a sense in which the requirements seem
more natural.  Within the set of initial conditions that experience sufficient inflation, all such states
give us reasonable universes at late times; in a more conventional Big Bang cosmology,
the perturbations require an additional substantial fine tuning.  Again, we have a 
relatively plausible target for a future comprehensive theory of initial conditions: as long as inflation
occurs, and the perturbations are not initially super-Planckian, we will get a 
reasonable universe.

These features of inflation are certainly not novel; it is well-known that inflation allows 
for the creation of a universe such as our own out of a small and relatively small bubble
of false vacuum energy.  We are nevertheless presenting the point in such detail because
we believe that the usual sales pitch for inflation is misleading; inflation does offer
important advantages over conventional Friedmann cosmologies, but not necessarily
the ones that are often advertised.  In particular, inflation does not by itself make our
current universe more likely; the number of trajectories that end up looking like our present
universe is unaffected by the possibility of inflation, and even when it is allowed only a 
tiny minority of solutions feature it.  Rather, inflation provides a specific kind of 
set-up for a true theory of initial conditions -- one that is yet to be definitively developed.

\section*{Acknowledgments}

This work was supported in part by the
U.S. Dept. of Energy and the Gordon and Betty Moore Foundation.  We thank Andy Albrecht,
Adrienne Erickcek, Don Page, and Paul Steinhardt for helpful conversations.

\section{Appendix: Eternal Inflation}

Eternal inflation \cite{Vilenkin:1983xq,Linde:1986fc,Linde:1986fd,Guth:2007ng}
is sometimes held up as a solution to the puzzle of the unlikeliness of inflation occurring.
In many models of inflation, the process is eternal -- while some regions reheat and
become radiation-dominated, other regions (increasing in physical volume) continue
to inflate.  This may be driven by the back-reaction of large quantum fluctuations in the 
inflaton during slow-roll inflation, or simply by the failure of percolation in a false vacuum
with a sufficiently small decay rate.

Through eternal inflation, a small initially inflating volume grows without bound, creating
an ever-increasing number of pocket universes that expand and cool in accordance with
conventional cosmology.  Therefore, the reasoning goes, it doesn't matter how unlikely
it is that inflation ever begins; as long as there is some nonzero chance that it starts, it
creates an infinite number of universes within the larger multiverse, and questions of 
probability become moot.

If unitary evolution is truly respected, this reasoning fails.  Consider the state of the 
universe at some late time $t_*$ (long after inflation has begun), in some particular slicing.  
Let us imagine that the basic idea of eternal inflation is correct, and the multiverse
consists of more and more localized universes of ever-increasing volume as time passes.
According to the reasoning developed in this paper, the macroscopic state of the multiverse
(that is, the set of microstates with macroscopic features identical to the multiverse
at time $t_*$) will be compatible with a very large number of past histories, only a very
small fraction of which will begin in a single inflating patch.  The more the volume grows
and the more universes that are created, the \emph{less} likely it is that this particular
configuration began with such a patch.  It requires more and more fine-tuning to take all
of the degrees of freedom and evolve them all backward into their vacuum states in a Planck-sized region.
Therefore, while eternal inflation can create an ever-larger volume, it does so at the expense
of starting in an ever-smaller fraction of the relevant phase space.

To say the same thing in a different way, if a multiverse mechanism is going to claim to solve
the cosmological fine-tuning problems, it will have to be the case that the mechanism
applies to generic (or at least relatively common) initial data.  We should be able to start
from a non-finely-tuned state, evolve it into the future (and the past), and see universes
such as our own arise.  As conventionally presented, models of eternal inflation usually
presume a starting point that is a smooth patch with a Planckian energy density -- very far
from a generic state.  If it could be shown that eternal inflation began from generic initial 
data, this objection would be largely overcome.  Presumably the resulting multiverse would
be time-symmetric on large scales, as in \cite{Aguirre:2003ck,Carroll:2004pn,Hartle:2008ng}.

It is possible that considering the entire multiverse along a single time slice is illegitimate,
and we should follow the philosophy of horizon complemenarity and only consider spacetime
patches that are observable by a single worldline.  This approach would run into severe
problems with Boltzmann brains if our current de~Sitter vacuum is long-lived
\cite{Dyson:2002pf,Albrecht:2004ke,Page:2006dt,Bousso:2006xc}.  Alternatively,
we might argue that the phase space is infinitely big,
and there is no sensible way to talk about probabilities.  That may ultimately be true, but represents
an abandonment of any hope of explaining cosmological fine-tuning via inflation, rather than
a defense of the strategy.  

Analogous concerns apply to cyclic cosmologies \cite{Steinhardt:2001st}.  Here, conditions
similar to our observable universe happen multiple times, separated primarily in time rather
than in space.  But the burden still remains to show that the conjectured evolution would proceed
from generic initial data.  The fact that the multiverse is not time-symmetric (the arrow of time
points in a consistent direction from cycle to cycle) makes this seem unlikely.


\begin{thebibliography}{999}
\parindent=.6em


\bibitem{Guth:1980zm}
  A.~H.~Guth,
  Phys.\ Rev.\  D {\bf 23}, 347 (1981).

\bibitem{Linde:1981mu}
  A.~D.~Linde,
  Phys.\ Lett.\  B {\bf 108}, 389 (1982).

\bibitem{Albrecht:1982wi}
  A.~Albrecht and P.~J.~Steinhardt,
  Phys.\ Rev.\ Lett.\  {\bf 48}, 1220 (1982).
  
  \bibitem{penrose}
  R. Penrose,
  in S. W. Hawking and W. Israel. 
  {\it General Relativity: An Einstein Centenary Survey}. Cambridge University Press. 
   pp. 581Ð638. (1979).
  
\bibitem{Gibbons:1986xk}
  G.~W.~Gibbons, S.~W.~Hawking and J.~M.~Stewart,
  Nucl.\ Phys.\  B {\bf 281}, 736 (1987).

\bibitem{Hawking:1987bi}
  S.~W.~Hawking and D.~N.~Page,
  Nucl.\ Phys.\  B {\bf 298}, 789 (1988).

\bibitem{Coule:1994gd}
  D.~H.~Coule,
  Class.\ Quant.\ Grav.\  {\bf 12}, 455 (1995)
  [arXiv:gr-qc/9408026].
  
\bibitem{Unruh:1996sf}
  W.~G.~Unruh,
{\it  In *Princeton 1996, Critical dialogues in cosmology* 249-264}.  

\bibitem{Hollands:2002yb}
  S.~Hollands and R.~M.~Wald,
  Gen.\ Rel.\ Grav.\  {\bf 34}, 2043 (2002)
  [arXiv:gr-qc/0205058].

\bibitem{Kofman:2002cj}
  L.~Kofman, A.~Linde and V.~F.~Mukhanov,
  JHEP {\bf 0210}, 057 (2002)
  [arXiv:hep-th/0206088].

\bibitem{Hollands:2002xi}
  S.~Hollands and R.~M.~Wald,
  arXiv:hep-th/0210001.

\bibitem{Carroll:2004pn}
  S.~M.~Carroll and J.~Chen,
  arXiv:hep-th/0410270.

\bibitem{Gibbons:2006pa}
  G.~W.~Gibbons and N.~Turok,
  arXiv:hep-th/0609095.
  
\bibitem{Linde:2007fr}
  A.~Linde,
  arXiv:0705.0164 [hep-th].
  
\bibitem{Vachaspati:1998dy}
  T.~Vachaspati and M.~Trodden,
  Phys.\ Rev.\  D {\bf 61}, 023502 (1999)
  [arXiv:gr-qc/9811037].

\bibitem{Mathur:2008ez}
  S.~D.~Mathur,
  J.\ Phys.\ Conf.\ Ser.\  {\bf 140}, 012009 (2008)
  [arXiv:0803.3727 [hep-th]].
    
\bibitem{Egan:2009yy}
  C.~A.~Egan and C.~H.~Lineweaver,
  Astrophys.\ J.\  {\bf 710}, 1825 (2010)
  [arXiv:0909.3983 [astro-ph.CO]].
  
\bibitem{Greene:2009tt}
  B.~Greene, K.~Hinterbichler, S.~Judes and M.~K.~Parikh,
  arXiv:0911.0693 [hep-th].

\bibitem{Garriga:1999vw}
  J.~Garriga and V.~F.~Mukhanov,
  Phys.\ Lett.\  B {\bf 458}, 219 (1999)
  [arXiv:hep-th/9904176].
  
\bibitem{Freese:1990rb}
  K.~Freese, J.~A.~Frieman and A.~V.~Olinto,
  Phys.\ Rev.\ Lett.\  {\bf 65}, 3233 (1990).

\bibitem{Freese:2008if}
  K.~Freese, C.~Savage and W.~H.~Kinney,
  Int.\ J.\ Mod.\ Phys.\  D {\bf 16}, 2573 (2008)
  [arXiv:0802.0227 [hep-ph]].

\bibitem{Mukhanov:1990me}
  V.~F.~Mukhanov, H.~A.~Feldman and R.~H.~Brandenberger,
  Phys.\ Rept.\  {\bf 215}, 203 (1992).

\bibitem{Mukhanovtext}
  V. F. Mukhanov, 
  \emph{Physical Foundations of Cosmology} (2005), Cambridge University Press.
  
\bibitem{Trodden:2004st}
  M.~Trodden and S.~M.~Carroll,
  arXiv:astro-ph/0401547.

\bibitem{Hartle:2008ng}
  J.~B.~Hartle, S.~W.~Hawking and T.~Hertog,
  Phys.\ Rev.\  D {\bf 77}, 123537 (2008)
  [arXiv:0803.1663 [hep-th]].

\bibitem{Vilenkin:1983xq}
  A.~Vilenkin,
  Phys.\ Rev.\  D {\bf 27}, 2848 (1983).

\bibitem{Linde:1986fc}
  A.~D.~Linde,
  Mod.\ Phys.\ Lett.\  A {\bf 1}, 81 (1986).

\bibitem{Linde:1986fd}
  A.~D.~Linde,
  Phys.\ Lett.\  B {\bf 175}, 395 (1986).

\bibitem{Guth:2007ng}
  A.~H.~Guth,
  J.\ Phys.\ A  {\bf 40}, 6811 (2007)
  [arXiv:hep-th/0702178].

\bibitem{Dyson:2002pf}
  L.~Dyson, M.~Kleban and L.~Susskind,
  JHEP {\bf 0210}, 011 (2002)
  [arXiv:hep-th/0208013].
  
\bibitem{Albrecht:2004ke}
  A.~Albrecht and L.~Sorbo,
  Phys.\ Rev.\  D {\bf 70}, 063528 (2004)
  [arXiv:hep-th/0405270].
  
\bibitem{Page:2006dt}
  D.~N.~Page,
  Phys.\ Rev.\  D {\bf 78}, 063535 (2008)
  [arXiv:hep-th/0610079].

\bibitem{Bousso:2006xc}
  R.~Bousso and B.~Freivogel,
  JHEP {\bf 0706}, 018 (2007)
  [arXiv:hep-th/0610132].
  
\bibitem{Aguirre:2003ck}
  A.~Aguirre and S.~Gratton,
  Phys.\ Rev.\  D {\bf 67}, 083515 (2003)
  [arXiv:gr-qc/0301042].
  
\bibitem{Steinhardt:2001st}
  P.~J.~Steinhardt and N.~Turok,
  Phys.\ Rev.\  D {\bf 65}, 126003 (2002)
  [arXiv:hep-th/0111098].

\end{thebibliography}
\end{document}